JACEK BIAŁOWĄS

# TOPOGRAPHY OF THE NUCLEI AND DISTRIBUTION OF ACETYLCHOLINESTERASE ACTIVITY IN THE SEPTUM OF THE TELENCEPHALON IN MAN

Departrnent of Normal Anatomy, Institute of Medical Biology, Medical Academy, Gdańsk
Director of the Department: Prof. Dr. O. Nar'kiewicz

Distribution of acetylcholinesterase (AChE) activity in the septum of the telencephalon in man was studied in 15 human brains using the acetylthiocholine method. Highest activity of AChE was found in the nucleus of the diagonal band and nucleus accumbens, and lowest in the Lateral nucleus. Comparison of histochemical results with cellular structure and wi.th the course of fibers showed absence in man of some of the nuclei described in animals such as the anterior medial nucleus, triangular nucleus, and marked reduction of the septo-hippocampal nucleus and fimbriate nucleus. Areas of the septum showing AChE activity were divided into an anterior and posterior system. The applicability to man of some neurophysiologic findings in animals is discussed.

The region of the septum of the telencephalon is one of the centers of the so-called limbic system, which embraces the cortex of the cingulated gyrus and some subcortical centers such as the amygdaloid body and hippocampus. The name of this septum should not be confused with that of the septum pellucidum, which applies only to the upper, thinned, part of the septum in man and some monkeys. On the other hand, it includes the much wider basal part and the paraterminal gyrus on the medial surface of the hemisphere, corresponding to the vertical part of the diagonal band of Broca. Studies on some animals show that this area contains, among others, inhibitory systems which are active in emotional states (6, 9, 10, 14, 15, 17, 28). Lesion of the septum is followed by increased activity in rats and cats, which become less timorous. *Zeeman* and *King* (30) observed similar symptoms in human beings with tumors in this area. *Nowicki* (22), in a pneumoencephalographic study, noted a relation between symptoms of characteropathy and thickening of the septum in man. Various vegetative disorders have been reported after lesions of the septum in rats (for literature, see *Cytawa*- 9).
*Hamilton* et al. and *Igic* et al. (15, 17) noted an effect of atropine and amiton on functions of the septum, suggesting existence of specific systems of cholinergic conduction in this area. Animals treated locally with these drugs exhibit symptom s suggesting damage of the septum. Studies on rats, cats and monkeys (4, 5, 12, 13, 18, 20, 21, 27) showed characteristic distribution of acetylcholinesterase (AChE) in the septal nuclei. We considered that a similar study in man would be interesting, especially since, according to the literature, cholinergic transmission seems to play an important role in this region (4, 5, 15, 17, 19, 20, 23).

## MATERIAL AND METHODS

Sections from 15 human brains obtained from the Department of Pathological Anatomy of the Medical Academy, Gdańsk, taken 19-30 hours after death from subjects aged 2 weeks to 74 years, were studied. A separate subgroup consisted of brains from children (1 from a 2-weekold infant, and 2 from 3-month-old children); all other brains were from adults. The patients from whom the brain sections were taken had died from causes unrelated to the central nervous system, with the exception of two who had hemorrhagic foci in the brain, but distant from the septum, which was grossly unaffected. After fixation in 10% formalin neutralized with calcium carbonate and cooled to 4°C, serial coryostat sections were stained by Koelle's acetylthiooholine method in the modification of Gerebtzoff (11). Etopropazine was added to the incubation mixture in the concentration of $10^{-4}M$ to inhibit nonspecific cholinesterases. The reaction was carried out at pH 5.6 at room temperature. Sections next to those stained with acetylthiocholine were stained with cresyl violet according to Nissl and with hematoxylin by the method of Weil.

## RESULTS

Basing on a preliminary analysis of preparations stained with cresyl violet and on published data from similar studies on animals (1, 2, 4, 5, 12, 13, 16, 21, 29), the septum was divided as proposed by *Andy* and *Stephan* (3) with some modifications. The general scheme of the different parts of the septum in man in serial frontal sections is illustrated in Fig. 1.

The septum contains the fallowing nuclei:
1. Septo-hippocampal nucleus (HA),
2. Nucleus of the vertical part of the diagonal tract of Broca (NB),
3. Nuc1eus accumbens of the septum (NA),
4. Lateral nuc1eus (L),
5. The dorsal nuc1eus, consisting of an inner (Dl), intermediate (DM) and outer part (DE),
6. The fimbriate nucleus (NF).

First, typical sections of the septum, from before backward, stained for AChE, will be discussed.

### Anterior part of the septum

Fig. 2A represents the most anterior segment of the septum, with distinctly standing out rostrurn of the corpus callosum. AChE activity is increased in the septum pellucidum 'in the form of two symmetric bands. Just below the rostrum of the corpus callosum, a band of increased reactivity can be seen, which corresponds to the rudimentary form of the septo-hippocampal nucleus in man. In preparations stained by the method of Nissl, dark cells similar to those encountered in the hippocampus can be seen in moderate numbers. Thin active bands running to the sides and downward connect this area with the deepest layer of cortex.

The next section in' the backward direction (Fig. 2B) shows high AChE activity in the most anterior upper part of the diagonal band of Broca. An anteromedial nuc1eus with increased enzyme activity, as seen in animals, cannot be recognized here.

Below, the dark area of the nucleus accumbens forming the lower limit of the ventric1e and indenting the septum is seen. In this region, high AChE activity has characteristic diffuse dark-brown coloration. The area of increased AChE activity coincides strictly with the boundaries of the .nucleus accumbens, as determined by cellular structure. The rather small cells intensely stained by cresyl violet with characteristic fine granules in their cytoplasm farm irregular groups, especially in the medial part of the nuc1eus accumbens. It is hard to say whether the increased reaction pertains mainly to neuropil, as also to the cells as suggested by different intensity of staining corresponding to the cell groups described above.

Between the nuc1eus accumbens and nuc1eus of the diagonal band, bands giving an intense 'reaction for AChE connect the two. They run diagonally upward and medial ward showing stratified activity; their external part merges with the lateral part of the septum, while only the internal part reaches the upper part of the diagonal tract. In preparations stained with cresyl violet, small groups of dark cells of a type similar to those in the nucleus accumbens can be seen. These areas are usually distinguished by darker brown coloration, indicating high AChE content. In horizontal sections (Fig. 4 C, D) these bands extend fairly far in the forward direction, running on the boundary between the deepest layer of the cortex and the white substance of the cingulum.

### Middle part of the septum

The next sections (Fig. 2C, D) represent the middle part of the septum. The distinctly dark region in the medial part of the septum in the neighborhood of the longitudinal fissure of the brain is the nucleus of the vertical part of the diagonal band of Broca. The boundaries of the area showing increased AChE activity in the form of diffuse brown coloration of varying intensity are quite sharp. They correspond to the area containing the large irregularly shaped cells with cresyl-violet stained granules in their cytoplasm, which are characteristic of the nucleus of the diagonal band. The fairly numerous darker spots on the brown background probably represent activity of the enzyme in the perikaryon of these cells; the nuclei, on the other hand, give a negative reaction. In sections further backward (Fig. 3A), the active area of the nucleus of the diagonal band is reduced to a thin layers surrounding bundles of fibers of the fornices from the side and from below, as can be seen also in the horizontal sections (Fig. 4A, B).

Laterally to the diagonal tract and above the nucleus accumbens, the basal area of the septum, occupied by the lateral nucleus, is characterized by lack of AChE activity (Fig. 2, B, C, D). In preparations stained with cresyl violet, the cells of this nucleus have various shape and size and are not very intensely stained so that the boundaries of the nucleus are not sharply defined. In preparations stained by the thiocholine method, the boundaries, particularly with the dorsal nucleus, are somewhat more distinct (Fig. 3A),

The next frontal section (Fig: 3A) visualizes the part of the septum where the crura of the anterior commissure approach the median plane and tend to join. Besides the previously mentioned thin band surrounding the columns of the fornix from the sides, which is the furthest backward reaching part of the diagonal band, two other thin bands of strong AChE activity are visible in the lateral and upper area of the septum, situated in the so-called dorsal nucleus of the septum, in its lateral part upward and backward of the lateral nucleus. In preparations stained with cresyl violet, three layers can be distinguished, although not as distinctly as in some animals:

J. Białowąs

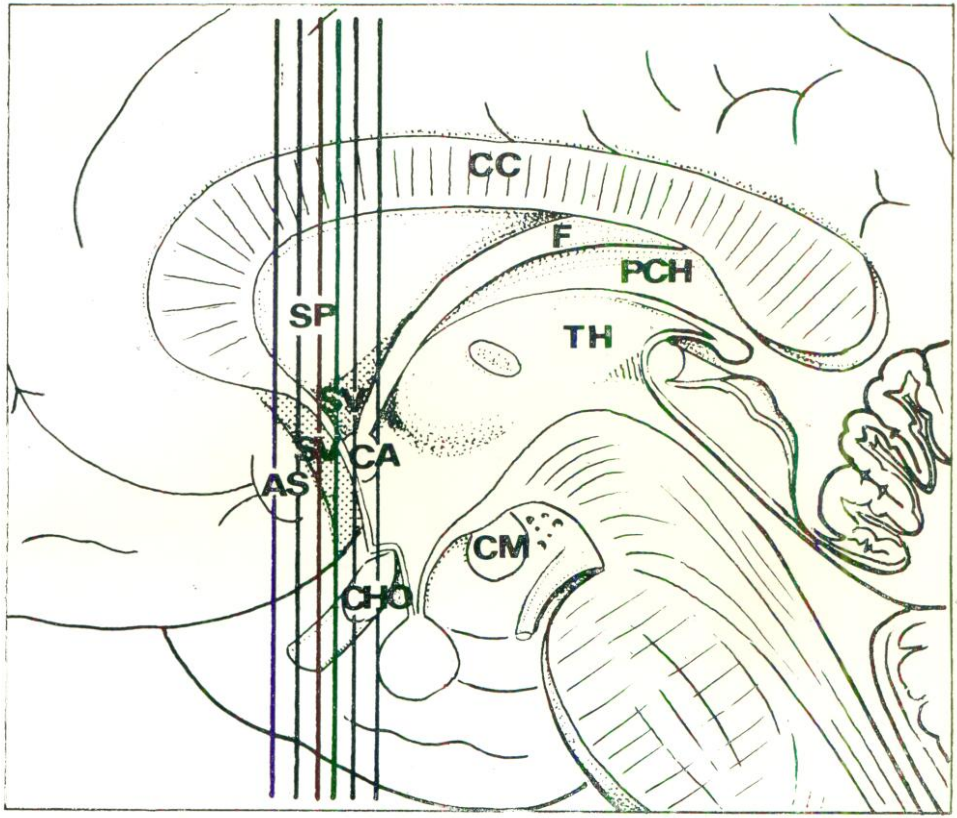

*Fig.* 1. Medial surface of the cerebral hemisphere in man. Planes of the frontal sections in Figs. 2 and 3 are shown.
Designations : AS - *area subcallosa,* CA - *commissura anterior,*
CC - *corpus callosum,* CHO - *chiasma opticum,* CM - *corpus mamillare,* F - *jornix,* PCH - *plexus chorioideus,* SP - *septum pellucidum,* SV - *septum verum,* TH - *thalamus.*

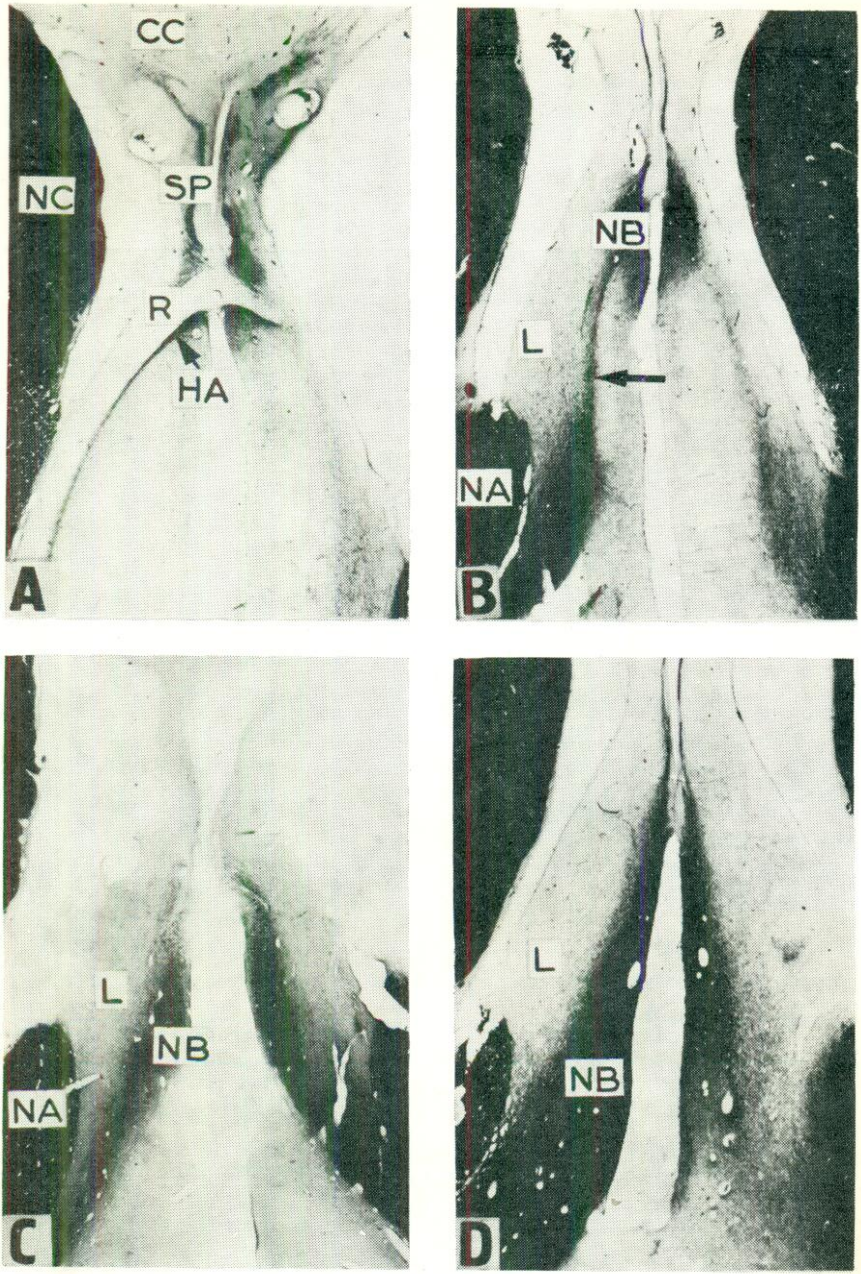

*Fig.* 2. Consecutive frontal sections through the Septum in man. Thiocholine technique. The nuclei are designated, A, B, D - magn, X6; C - magn. X4. The arrow points to active bands connecting the nucleus accumbens with the nucleus of the diagonal band.
Designations : CA - *commissura anterior,* CC - *corpus callosum,*
HA - *nucl. septo-hippocampalis,* L - *nucl. lateralis,* NA - *nucl. accumbens,* NB - *nucl. tractus diagonalis Broca,* NC - *nucl. caudatus,* R - *rostrum corporis callosi,* SP - *septum pellucidum.*

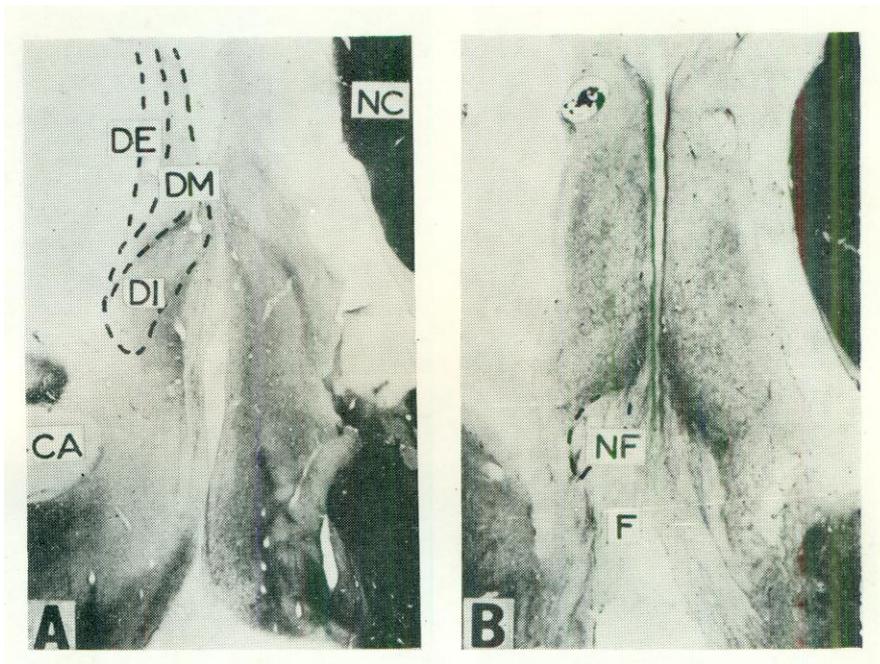

*Fig.* 3. Further frontal sections through the septum in man. Thiocholine technique.
A - magn, X4; B - magn, X6.
Designations : CA - *commissura anterior,* DE - *nucl. dorsalis - pars externa,* Dl - *nucl. dorsalis - pars interna,* DM - *nucl. dorsalis - pars intermedia,* F - *fornix,* NC - *nucl. caudatus,* NF - *nucl. fimbratus septi .*

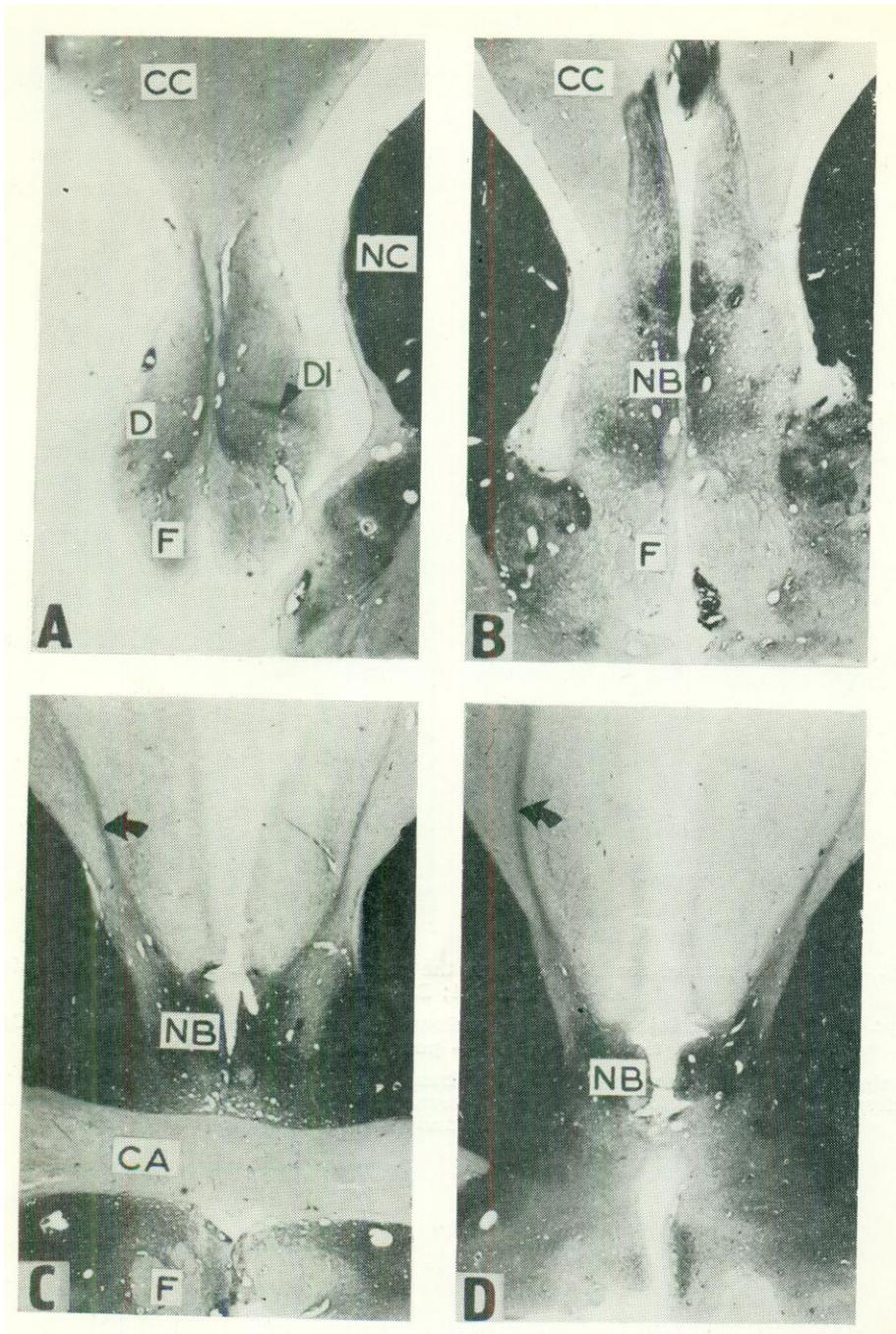

*Fig.* 4. Horizontal sections through the septum in man, from above downward. Thiocholine technique. A, B, C, D - magn. X3. Arrows point to AOhE activity of the band s connecting the septum with the cortex. Designations as in the preceding figures (Dorsal nucleus – D).

: an internal layer, neighboring with the posterior part of the diagonal band and fornix, and containing fairly intensely stained cells; an intermediate part, and an external part, neighboring with the ventricle, with less densely distributed cells. The course of the bands of intense AChE activity agrees with the course of the layers, but the bands are narrower than the layers. The first band, reaching furthest downward, is situated in the internal part; and the second band, which is very poorly visible, is situated in the intermediate part. A stratification of the reaction in the postero-superior part of the septum can be discerned also in the horizontal sections (Fig. 4A).

### Posterior part of the septum

The last of the frontal sections (Fig. 3B) represents the septum just before the interventricular foramen. The upper part of the septum is thick and hardly fits here the term septum pellucidum any more. The columns of the fornix are inactive, and higher up the somewhat darker area of the dorsal nucleus can be seen. Of the bands of increased activity, only the internal one, separating the fimbriate from the dorsal nucleus, remains. Compared to other animals, the area occupied by the fimbriate nucleus is small. Only a few dark bands, presumably representing acetylcholinesterase-rich nerve fibers, can be seen above the columns of the fornix.

The triangular nucleus below the columns is indistinguishable in man, probably owing to marked reduction of the anterior commissure of the fornix to a thin band of fibers connecting both fornices on the ventral side, posteriorly to the interventricular foramen. No significant differences were noted between the preparations incubated with etopropazine and those incubated without it. Preparations from brains received 26-30 hours after death gave less strongly positive reactions compared with fresher brains.

### DISCUSSION

The available literature contains precise descriptions of the distribution of acetylcholinesterase activity in the septum in the cat, monkey and rat (4, 5, 12, 13, 16, 18, 21), but only fragmentary information about its distribution in man (16). The reports agree that highest AChE activity is found in the nuclei of the diagonal band and nucleus accumbens of the septum. According to the results of the present study, this applies also to man. In preceding studies (4, 5, 16, 21), stratification of the reaction in the dorsal nucleus in the cat and rat was described. *Girgiss (12,* 13) has described similar differentiation of the activity in *Galago* and monkeys. In the present study, layers of increased activity were observed in the area of the dorsal nucleus in man, in confirmation of the division of this nucleus into layers proposed by *Andy* and *Stephan* (1-3, 29). Similar results were found by *Harkmark* et al. (16).

In the light of the presented data, the division of *Andy* and *Stephan* reflects the structural organization better than the former division by *Shimazono* and *Brockhaus* (7, 8, 26). Nevertheless, AChE activity allows further simplification. For instance, an anterior medial nucleus, which in animals shows moderately high AChE activity, is not distinguishable in man. The area adjacent to the ventricle given this name by *Andy* and *Stephan* should be considered a part of the lateral nucleus because of the lack of AChE activity. The triangular nucleus is absent in man, and the septo-hippocampal and fimbriate nuclei are markedly reduced.

In previous studies on rats the acetylcholinesterase-rich centers in the septum were divided into connected systems. The same can be done with the human brain. For instance, the nucleus of the diagonal band together with the fimbriate nucleus form a single system which, according to experimental studies on animals, has massive cholinergic connections with the hippocampus (5, 19-21, 23-25), and which may be called the posterior system.

A second group of AChE-rich centers in the septum consists of the septo-hippocampal and medial anterior nuclei together with active bands which connect them with the cingulate cortex. In man, in spite of absence of the medial anterior nucleus, this system is better developed than in animals owing to the active bands reaching far into the cingulated cortex. The name anterior system is proposed.

The remaining centers showi.ng AChE activity are difficult to systematize. The nucleus accumbens is probably connected with the striatum, and the dorsal nucleus, which receives stratified projection from the hippocampus (24, 25), may also be the site of terminals of collaterals of fibers from the diagonal tract (16).

The studies cited in the introduction indicate that the septum plays a considerable role in the regulation of emotional states and vegetative functions in animals and in man. Establishment of homology of the various areas of the septum in animals and man by further comparative anatomical studies should permit comparison of the results of some experimental studies on animals with clinical observations.

The author's present contact address: jacekwb@gumed.edu.pl  (Dr. med. Jacek Białowąs in Polish , Jacek Bialowas MD in English),